\documentstyle[12pt]{article}
\newcommand{\be}{\begin{equation}}
\newcommand{\ee}{\end{equation}}
\newcommand{\bea}{\begin{eqnarray}}
\newcommand{\eea}{\end{eqnarray}}
\newcommand{\etal}{{\em et al.}}

\begin{document}
\newpage

\begin{center}
{\large\bf On signal-noise decomposition of timeseries using the continuous wavelet transform: Application to sunspot index}

\vspace{.5cm}

J. Polygiannakis,$^1$ P. Preka-Papadema $^1$ and X. Moussas $^1$

\vspace{.5cm}

{\em 1. Section of Astromony, Astrophysics and Mechanics, Department of
Physics,\\ National University of Athens, GR 15784, Athens, Greece. 
\\ e-mail: ipolig@cc.uoa.gr \\}

\end{center}

\vspace{1cm}

\small

{\bf Abstract.} We show that the continuous wavelet transform can provide a unique
decomposition of a timeseries in to 'signal-like' and 'noise-like'
components: From the overall wavelet spectrum two mutually independent
{\em skeleton spectra} can be extracted, allowing the separate detection
and monitoring in even non-stationary timeseries of the evolution of (a)
both stable but also transient, evolving periodicities, such as the
output of low dimensional dynamical systems and (b) scale-invariant
structures, such as discontinuities, self-similar structures or noise.
The idea of the method is to keep from the overall wavelet expansion of
the timeseries only the wavelet components of largest amplitude at any
given time or scale, thus obtaining the {\em instantly maximal} and {\em
scale maximal} wavelet skeleton spectrum, respectively. The scale
maximal spectrum was previously proposed for studying possible
multifractal scaling properties of the timeseries (e.g. Arneodo \etal,
1988). The here proposed instantly maximal spectrum exhibits clearer
spectral peaks and reduced noise, as compared to the overall wavelet
spectrum. An indicative application to the monthly-averaged sunspot
index reveals, apart from the well-known 11-year periodicity, 3 of its
harmonics, the 2-year periodicity ({\em quasi-biennial oscillation,
QBO}) and several more (some of which detected previously in various
solar, earth-solar connection and climate indices), here proposed being
just harmonics of the QBO, in all supporting the {\em double-cycle}
solar magnetic dynamo model (Benevolenskaya, 1998, 2000). The scale
maximal spectrum reveals the presence of $1/f$ fluctuations with
timescales up to 1 year in the sunspot number, indicating that the solar
magnetic configurations involved in the transient solar activity
phenomena with those characteristic timescales are in a
self-organized-critical state (SOC), as previously proposed for the
solar flare occurence (Lu \& Hamilton, 1991).

\normalsize

\section{Introduction}

The widely used Fourier transform, although useful for stationary
signals of simple dynamics consisted of a linear superposition of few
independent, strong, non-evolving periodicities, has severe drawbacks
for analyzing signals of the following two important categories, often
encountered as output of physically interesting complex systems:

a. Signals that include transient or variable periodicities. The Fourier
transform, since integrating over the whole time domain, does not allow
monitoring of amplitude or frequency evolution in periodicities. A
limited solution is to divide the signal in segments or else 'windows'
(a method called short-time, Gabor or windowed Fourier transform),
however introducing the arbitrary, fixed window width that imposes a low
frequency cut-off near which edge effects occur, thus artificially
deforming the estimated spectrum.

b. Signals that significantly depart from stationarity, consisted of
intermittent 'activity bursts' (e.g. energy releases), or hierarchies of
localized structures in space or time (e.g. in turbulent fluids reside
eddies, discontinuities, filaments, sheets, shocks etc.). For each
localized structure a wide frequency range of strongly phase coherent
(i.e. 'synchronized' in time) Fourier sine components are required to
reproduce it, adding near and canceling away from the structure. Hence,
each structure spreads and become undetectable over the whole Fourier
spectrum. Moreover, such structures often exhibit 'self-similarity', or
else 'scale invariance', i.e. consists of hierarchies of mutually
similar structures, emerging through a cascade mechanism, and
'roughness' over a wide scale range (well modeled by fractals and
multi-fractals) that is not built in the sine waves of Fourier
transform.

The projection of signals with continuous or discrete wavelet analyzing
functions, that are by definition self-similar and well localized in
{\em both} time (or space) and frequency (or wavenumber) domains,
provides efficient descriptions, in terms of sparseness, of
non-stationary, transient, or 'bursty' records. It is proven a useful
tool, used in an increasingly large variety of applications. Here we
mention some of them indicatively according to the type of use:
detection of periodic signals in noisy timeseries (e.g. Otazu \etal,
2002), monitoring of period variations (e.g. Frick \etal, 1997, Fligge
\etal, 1999), unevenly sampled period analysis (e.g. Foster, 1996),
derivation of multifractal properties (e.g. Arneodo \etal, 1988, Argoul
\etal, 1989, Muzy \etal, 1991, 1993, Arneodo \etal, 1995, 1998), random
multifractal synthesis (e.g. Benzi etal, 1993), localized structure
identification (e.g. Lucek \& Balogh, 1997, Roux \etal, 1999, Mouri
\etal, 1999), polarization analysis (e.g. Baumjohann \etal, 1999),
denoising (e.g. Fligge \& Solanki, 1997, Komm \etal, 1999),
signal-to-noise ratio enhancement (e.g. Zhang \& Paulson, 1997),
information compression (e.g. Muhlmann \& Hanslmeier, 1996),
multi-resolution image decomposition (e.g. Mallat, 1989, Pantin \&
Starck, 1996), studies of random walks (Arneodo \etal, 1996), flow
structure analysis (Haynes \& Norton, 1993), population classification
(Bendjoya \etal, 1991), clustering detection (e.g. Bijaoui \etal, 1993,
Girardi \etal, 1997, Lima Neto \etal, 1997), fast data query (e.g.
Chakrabarti \etal, 2001), fluid turbulence simulations (e.g. Schneider
\etal, 1997) etc.

The wavelet transform is now used extensively for the analysis of
oscillations and periodicities in various solar structural and activity
features, as the sunspot number (Fligge \etal, 1999, Mordvidov \&
Kuklin, 1999, Sello, 2000, Rigozo \etal, 2001, Nayar \etal, 2002),
sunspot groups (Ballester \& Oliver, 1999), active regions (Ireland
\etal, 1999, O'Shea \etal, 2001, 2002), coronal holes (Marsh \etal,
2002), bursts (Schwarz \etal, 1998, Meszarosova \etal, 1999, Gimenez
\etal, 2001), flares (Aschwanden \etal, 1998), polar plumes (Banerjee
\etal, 2000), coronal loops (De Moortel \etal, 2000, 2002a, b),
photospheric flows (Lawrence \etal, 2001), chromosphere (Bocchialini \&
Baudin, 1995, Frick \etal, 1997), transition region (Fludra, 2001),
differential rotation (Hempelmann \& Donahue, 1997, Soon \etal, 1999,
Hempelmann, 2002) etc.

In this article we demonstrate that two exactly defined, mutually
independent, complementary wavelet skeleton spectra, extracted from the
total continuous wavelet spectrum, allow to discriminate the components
of (stable or transient) periodicities and of hierarchies of
discontinuities in timeseries. An indicative application is made to
sunspot number, since (a) is known to include both periodic and
noise-like components, and (b) its study may contribute in understanding
the features and mechanism of the solar activity. Section 2 includes a
brief review of the continuous wavelet transform for the unfamiliar
reader (for extended reviews see e.g. Hunt \etal, 1993, Kaiser, 1994,
Louis \etal, 1997, Torrence \& Compo 1998). The two wavelet skeleton
spectra are introduced in Section 3. The application to the sunspot
index is presented in Section 4. Conclusions and discussion are
summarized in Section 5.

\section{A brief review of the wavelet transform}

\subsection{Linear functional projection}
\label{RedudancyComment}

One way to quantify the degree of 'similarity' between two, generally
complex function $x(t),\; y(t)$ of a variable $t \in [0,T]$ (e.g. time
or distance) is by the amplitude (or else cross-correlation) of their
{\em functional projection} over the domain of $t$:

\be
a= \frac{1}{T} \int_{0}^{T} x(t) \; y^*(t) \; dt
\ee
with $y^*$ being the complex conjugate of $y$. The factor $1/T$ makes
the amplitude $a$ independent of $T$ and finite in the limit $T
\rightarrow +\infty$. The two functions are called 'orthogonal' if
$a=0$.

In analogy to the projection of a vector to a coordinate system in
physical space, a given function $x(t)$ (e.g. the timeseries of a
measured observable) thus can be analyzed to a family of functions
$y(t,p_i)$, of $n$ characteristic parameters $p_i, \; i=1...n$:

\be
a(p_i)= \frac{1}{T} \int_{0}^{T} x(t) \; y^*(t,p_i) \; dt
\label{projection}
\ee
The functional analysis $(\ref{projection})$ is linear in the sense that
any $x(t)$ can be expanded to a {\em linear} superposition of a set of
functions $y(t,p_i)$ that are mutually orthogonal and integrable.

Note that the continuous functional transform $(\ref{projection})$ is
generally {\em redundant} since it spreads the information content of
$x(t)$ from the one-dimensional time axis, to the n-dimensional space of
the parameters $p_i$. The minimization of information spread in the
parameter space $p_i$ by appropriate selection of the functional basis
and the development of efficient redundancy reduction schemes is of
central importance for spectral analysis, functional approximation and
compression applications.

\subsection{Appropriate selection of analyzing functions}

Although the linear functional projection $(\ref{projection})$ is
mathematically valid with any set of functions, an {\em appropriate}
projection should aim in selecting such functions to match as much as
possible (i.e. being nearly orthogonal to most of the functions so
that the representation is sparse) the analyzed function, which often is
or consists of the following:

(a) {\em Analog signal}, as the independent variables of a system or
model of relatively simple dynamics. The usually smoothly varying
variables of systems of low dimensionality (i.e. small number of
independent variables) exhibit oscillatory behavior, thus can often be
sparsely expanded to a linear superposition of a few periodic functions,
or, more generally, transient, quasi-periodic components. Hence, it is
desirable that the analyzing functions have well-defined, single-peaked,
narrow frequency spectral content.

(b) {\em Noise}, as this due to observational 'random' fluctuations
(often additive) of high-frequency content, e.g. of flat spectrum in
uncorrelated (white) or more generally of wide range power-law decaying
spectrum in auto-correlated ('colored') noise. Noise-like contributions
may also caused by fast-varying components of a deterministic complex
system of large number of independent variables (such as macroscopic
systems) or the interaction of a low-dimensional open system with an
environment of high dimensionality. The observables of high-dimensional
systems, often involving strongly turbulent fluids, are characterized by
intermittency or 'singularities' due to coherent structures or
intermittent energy bursts and, exact or weaker, {\em self-similarity}
between different scales (often termed {\em scale-invariance}), well
modeled by fractals and multi-fractals. Although such systems are
characterized by strongly non-linear physical mechanisms, for an
appropriate {\em linear} projection of such observables, the analyzing
functions should at least posses the element of self-similarity.

\subsection{Fourier transform}
\label{FourierTransform}

The widely used Fourier transform can be interpreted as {\em mapping},
or complete 'rotation' (in analogy to the vector rotation in physical
space) from the time domain to the frequency domain, by means of the
complex periodic plain wave functions:

\be
y(t,f) = e^{2\pi i f t}
\label{FourierBasis}
\ee
The (generally complex) {\em spectral amplitude} (see equ.
$(\ref{projection})$):

\be
a(f) = \frac{1}{T} \int_{0}^{T} x(t) \; y^*(t,f) \; dt
\label{FourierAmpl}
\ee
thus represents an estimate of the amplitude of periodicities of
frequency $f$, that occurred during the time interval $T$. The, often
studied, Fourier power spectral density is defined as follows:

\be
p(f)=a^2(f)
\label{FourierPSD}
\ee
A useful property of the Fourier transforms is the {\em power theorem}:
any linear functional projection $(\ref{projection})$ can be calculated
via Fourier transforms in the frequency domain (e.g. Hunt \etal, 1993, p.6):

\be
a(p_i)= \int_{0}^{T} x(t) \; y^*(t,p_i) \; dt
= \int_{-\infty}^{+\infty} \hat{x}(f) \; \hat{y}^*(f,p_i) \; df
\label{powertheorem}
\ee
where $\hat{x},\hat{y}$ the Fourier transforms of $x,y$ respectively
(see $(\ref{FourierAmpl})$). Hence, the functional projection
$(\ref{projection})$ can be treated as a {\em linear filter} acting on
$\hat{x}$.

\subsection{Time-frequency analysis}

Note that each Fourier basis function $y$ is totally localized in
frequency but nowhere in time, since having infinite duration (i.e. $y$
in $(\ref{FourierBasis})$ do not vanish for $t \rightarrow \pm \infty$).
As a consequence, the Fourier transform does not allow monitoring of
transient (periodic or aperiodic) components in the signal, since they
are spread over the whole Fourier spectrum.

In the framework of the transform $(\ref{projection})$ a solution to
this problem is to introduce analysing functions of well-defined frequency
but are also localized in time i.e. of significant amplitude for a
finite time duration.

Perfect localization in both time and frequency in however inhibited by
the {\em uncertainty principle} (e.g. Kaiser, 1994, p. 50):
$(\ref{projection})$ keeps the information of $x(t)$ confined in {\em
cells}, within which there is total uncertainty about the information of
$x(t)$, and the size of these cells cannot be infinitesimally small,
having a lower limit depending on the properties (shape, rate of decay
etc.) of the selected analysing functions.

Any time-frequency transform represents a mapping of $x(t)$ on the $t-f$
plane. Given a set of (generally complex-valued) functions $y(t,t',f)$,
of characteristic frequency $f$, localized in the vicinity of time
$t'$, the linear functional projection $(\ref{projection})$ is:

\be
a(t,f)= \frac{1}{T} \int_{0}^{T} x(t') \; y^*(t,t',f) \; dt'
\label{CWT}
\ee
The graph of the power $a^2(t,f)$ on the $t-f$ plane is oftenly called
{\em scalogram}. The total spectrum:

\be
p(f) = \int_{0}^{T} a^2(t,f) \; dt
\ee
is expected to generally exhibit similar features and shape (e.g.
spectral peaks, slope) as the corresponding Fourier spectrum (equ.
$(\ref{FourierPSD})$).

\subsection{Windowed Fourier transform}

In the {\em windowed Fourier transform} (Gabor, 1946) the Fourier plain
wave functions are modulated by a 'support' $\phi$ (often called
'window' function) of significant value only in the vicinity $\sigma$ of
time $t'$:

\be
y(t,t',f) = e^{2\pi i f t} \phi(\frac{t-t'}{\sigma})
\label{gabor1}
\ee
Its Fourier transform has spectral width of the order of $1 / \sigma$.
Therefore, the $t-f$ plane is resolved in cells of fixed shape and area at
least $\sigma$ by $1/\sigma$.

A commonly used support function is a Gaussian of width $\sigma$:

\be
\phi(\frac{t-t'}{\sigma}) = e^{-\frac{(t-t')^2}{2 \sigma^2}}
\label{gabor2}
\ee
The resulting analysing function $(\ref{gabor1})$ has Fourier transform:

\be
\hat{y}(f,f') = \sigma e ^{\frac{-(f-f')^2 \sigma ^2}{2}}
\ee
The Gaussian support can be proven to give optimal localization in the
sense that among all functions it covers smallest area permitted by the
uncertainty principle in the time-frequency plane (Louis \etal, 1997).
Other window functions are also used, aimed at reducing the leakage of
power of spectral peaks to nearby frequencies (e.g. Press \etal, 1992).

However the windowed Fourier transform always introduces the {\em
arbitrary} duration of the support function ($\sigma $ in equ.
$(\ref{gabor2})$) which itself does not depend on $f$ and must be
selected appropriately before the analysis. Also, the analyzing
functions $y$ {\em are not self-similar} since include increasingly
many oscillations within their support with increasing frequency.

\subsection{Continuous wavelet transform}

The continuous wavelet transform resolves the information on the $t-f$
plane in cells of variable shape depending on frequency, with size of
the order of $1/f$ by $f$. With this partition of the $t-f$ plane the
arbitrary selection of the windowed Fourier transform is avoided and the
analyzing functions ({\em wavelets}) are exactly self-similar (or else
scale-invariant) since each includes a constant number of oscillations
independent of frequency. The wavelet transform at lower frequencies
provides better resolution in frequency but worse in time (since the
analysing functions are wider) and at high frequencies better time
localization (since the analysing functions are narrower) but more
uncertainty in frequency, in all acting as a 'mathematical microscope'.

While many wavelet families are so far proposed (also depending on the
nature of application), of the most commonly used in physical
applications is the {\em Morlet wavelet} (Morlet \etal, 1982), being the
generalization of the windowed Fourier transform (see equ.
$(\ref{gabor1})$):

\be
y(t,t',f) = e^{2\pi i f t} \phi(f(t-t'))
\label{gabor1wavelet}
\ee
with Gaussian support (in analogy with $(\ref{gabor2})$) of width $1/f$:

\be
\phi(f(t-t'))= e^{\frac{- f^2(t-t')^2}{2}}
\label{gabor2wavelet}
\ee
Its Fourier transform is also Gaussian of width $f$:

\be
\hat{y}(f,f') = \frac{1}{f} e ^{- \frac{(f-f')^2}{2 f^2}}
\ee
Among all possible wavelet analysing functions, the Morlet wavelet due
to its Gaussian support inherits optimality as regarding the uncertainty
principle (Louis \etal, 1997).

\section{The two wavelet skeleton spectra}

\subsection{Signal-noise decomposition:}

Successful decomposition of the timeseries into 'signal' and noise-like
component is of fundamental importance in various applications, for the
separate study of the signal and noise properties (e.g. locating
spectral peaks and deriving scaling laws), denoising, compression etc.
The wavelet transform is promising for that decomposition, since
allowing monitoring of the evolution of time-localized wavepackets of
specific frequency, hence also of stable or transient periodicities and
noise-like spikes.

The overall wavelet scalogram generally contains a mixture of the signal
and noise characteristics. Imposing some thresholding logic (e.g. hard
or soft level, trend etc) to the wavelet components (i.e. a type of
wavelet band-pass filter) is expected to have limited success in the
case of strong, auto-correlated noise and/or weak, transient
periodicities. The deeper reason that such filters have limited success
in those cases is that the selection of the wavelet components that are
attributed to the signal or the noise is to some extend arbitrary, not
adaptively taking into account the information content in the original
timeseries.

As commented in section $(\ref{RedudancyComment})$ for any continuous
functional projection, the continuous wavelet transform is always {\em
redundant} in both time and frequency, since mapping the one dimensional
information in $x(t)$ on the (two dimensional) $t-f$ plane. It contains
correlations between the wavelet components that do not exist in the
timeseries but is significant at those (small) regions and scales where
the wavelet functions cross-correlation is also significant. As a
consequence, the higher the achieved resolution (i.e. sampling in
frequency), the higher the {\em smoothness} of the resulting spectrogram
due to the increased cross correlations between wavelet amplitudes at
nearby frequencies. The presence of noise-like components, e.g. due to
observational noise, contaminates ('blurs') the scalogram, by making
more uncorrelated wavelet amplitudes of nearby times and introducing
ridges across wide ranges of frequencies.

In contrast to the continuous, the {\em discrete} wavelet transform (the
review of which is outside the scope of this article) leads to
uncorrelated wavelet amplitudes if the discrete wavelet analysing functions
are orthogonal. However the lack or redundancy is not an advantage for
the signal-noise decomposition, since there is no profound criterion for
attributing the wavelet components to signal or to noise. Some
thresholding criterion can be imposed, as commonly used in signal
compression applications, again having a degree of arbitrariness.

From the above discussion emerges a plan for an efficient signal-noise
decomposition using the continuous wavelet transform, consisted of the
following steps:

(a) Continuous wavelet transform of the timeseries with sufficiently
high frequency resolution (sampling), thus producing a highly redundant
wavelet scalogram.

(b) A non-parametric, information dependent selection criterion that
disposes effectively the redundancy, in a way that selects separately
those wavelet components that can be with high probability attributed to
signal and to those to noise.

\subsection{Scale maximal wavelet skeleton spectrum}

A transient structure or burst of finite, characteristic duration
$\tau$, occurring (i.e. having center of mass) in the timeseries at time
$t$ is expected to cause a local increase of the amplitudes $|a(t,f)|$
of the timeseries wavelet transform in the vicinity of time $t$, at
wavelet period of the order of $\tau$ (or else frequency $1/\tau$),
while can extent across different timescales, e.g. a discontinuity
($\tau=0$) is wavelet-transformed to a superposition of wavelets with
the same center of mass at time $t$, at all timescales (and
frequencies).

For the study of such structures the {\em scale maximal wavelet skeleton
spectrum} is extracted from the overall wavelet spectrum, keeping only
those wavelet components of which the complex amplitude is {\em locally
maximum} across time at any given time-scale, i.e. those for which:

\be
\frac{\partial |a(t,f)|}{\partial t} = 0, \;\; \frac{\partial^2 |a(t,f)|}{\partial t^2} < 0
\label{sceleton1}
\ee
It is already recognized that the so defined spectrum (otherwise called
wavelet transform modulus maxima) contains all the important information
about the existence of hierarchies of discontinuities, usually appearing
as continuous lines across scales pointing at the time of the
discontinuity, hence can be used for detection of singularities (e.g.
Mallat, 1999, p. 176), or even approximate reconstruction of timeseries
(e.g. Mallat, 1999, p. 185 and references therein). It is also used for
the study scaling (e.g. possible multifractal) properties in timeseries:
the lines of each singularity is often, at least at high frequencies,
consisted of wavelet amplitudes that follow power-law dependence on
frequency across many orders of frequency range (i.e. for these ranges
there is no characteristic scale, or else there is some {\em
scale-invariance} in the timeseries), well modeled by {\em
multifractals}: If the timeseries is a realization of a multifractal
then the wavelet amplitudes of the locally scale-invariant skeleton
spectrum, at least at high frequencies, $f \rightarrow +\infty$, follow
power-law of the form (for a review see e.g. Arneodo \etal, 1995):

\be
|a(t,f)| \propto f^{-h(t)}
\label{Hurst}
\ee
where $h(t)$ is the {\em H$\ddot{o}$lder} (or else {\em local Hurst})
exponent, being a local measure of 'burstiness' in the timeseries at
time $t$ (for details see e.g. Muzy \etal, 1993).

Power-law spectral power dependance over frequency, at least over
several orders of frequency magnitude and especially tails at high
frequencies (or small scales) are fairly common in all physical
observables. In the special case of {\em uni-fractals} $h(t)$ has the
same value for all $t$. Of specific interest is the case of $1/f$ {\em
noise} (i.e. with $h=0$), constituting the hallmark of {\em
self-organized criticality}, (SOC, Bak \etal, 1987) i.e. the meta-stable
state close to a critical point of a universal class of systems
characterized by intermittency, well modeling aspects of solar flares,
earthquakes, geological formation, biological evolution, economy etc.
(for a review see Bak, 1996).

\subsection{Instantly maximal wavelet skeleton spectrum}

A transient or stable periodic component of instant period $T$ at time
$t$ in the timeseries is expected to cause increase of the wavelet
amplitudes $|a(t,f)|$ in the vicinity of wavelet period of the order of
$T$ (or else frequency $1/T$), while may extend across time as far as
its period $T$ remains constant in time.

For the detection of periodicities in timeseries here we introduce the
{\em instantly maximal wavelet skeleton spectrum}, extracted from the
overall wavelet spectrum, if keeping only those wavelet components that
are locally of maximum amplitude at a given time, i.e. only those for
which:

\be \frac{\partial |a(t,f)|}{\partial f} = 0, \;\; \frac{\partial^2
|a(t,f)|}{\partial f^2} < 0 \label{sceleton2} \ee From the above
discussion becomes clear that the so defined spectrum can be used for
detecting transient periodicities of varying period, and of course those
of stable period in the timeseries. Note that the instantly maximal
skeleton spectrum excludes the wavelet components that instantly extend
across time scales, thus it is less contaminated by the noise-like
scale-invariant component of the timeseries. On the other hand, the
scale maximal spectrum includes only those wavelet components that
extend across time scales, thus the scaling properties derived from it
are not distorted by possible transient periodicities in the timeseries
due to the presence of components of analog signal type. In that sense
the two skeleton spectra are complementary, permitting the separate
study of the noise-like and signal-like timeseries component properties
separately. Noise-like components of an hierarchy of discontinuities
also affects strongly the complex phases, $arg(a(t,f))$, of the wavelet
spectrum since the wavelet expansion requires 'synchronization' of
wavelets (i.e. to be placed nearby in time). For this reason only the
wavelet amplitudes are considered in $(\ref{sceleton2})$. However,
auto-correlated random noise has significant slowly varying components,
i.e. non-periodic trends at various scales (and in that sense the two
skeleton spectra do {\em not} completely separate signal from noise),
nevertheless introducing a power-law background without concentration of
power in spectral peaks. In all, the instantly maximal skeleton spectrum
is expected to exhibit more clearly the periodicities of a signal-like
component of timeseries, practically excluding uncorrelated hierarchies
of discontinuities due to noise.

\section{Application to the sunspot index}

\subsection{A review on solar activity and sunspots}

The solar activity (e.g. sunspots, flares, coronal mass ejections etc)
is rich, involving complex, not well understood magnetofluid processes,
extending in a wide range of spatial scales and durations. The sunspot
number is widely regarded as an easily collectable quantitive solar
activity measure. Sporadic, naked-eye observations exist in Chinese
dynastic histories since 28 BC, while systematic sunspot observations
are kept since the discovery of telescope in $1610$, while not uniformly
reliable (Waldmeier 1961, Eddy, 1977, Sonett 1983).

The sunspot number is dominated by the 11-year periodicity (well-known
Schwabe in 1843) connected to the 22-year cycle of the overall solar
magnetic field. The 11-year cycle is stable, exhibiting only small
variations of each cycle duration, as also shown using wavelet analysis
of sunspot number (Fligge \etal, 1999). However, the maximum number of
sunspots in each cycle varies significantly and quite unpredictably,
even nearly, but not totally, vanishing for many cycles. In all, the
sunspot record includes enigmatic epochs of suppression, as the {\em
Maunder Minimum} from 1645 to 1715 (e.g. Ribes \& Nesme-Ribes, 1993) and
{\em Sp$\ddot{o}$rer Minimum} in the $15^{th}$ century, as well as
epochs of enhanced activity, as the nowadays {\em Modern Maximum} and
{\em Medieval Maximum} in the $12^{th}$ century. The reason for the
large variations of the cycle's amplitudes and the epochs of suppressed
activity of Maunder minimum type is unclear, e.g. proposed being due to
chaotic behaviour of the nonlinear dynamo equations (e.g. Ruzmaikin,
1983, K$\ddot{u}$ker \etal, 1999) or due to stochastic instabilities
forcing the solar dynamo leading to on-off intermittency (e.g. Hoyng,
1988, 1993, Ossendrijver \& Hoyng, 1996, Schmitt \etal, 1996).

The average shape of each sunspot cycle is systematically asymmetric,
taking less time rising to maximum than reaching the next minimum,
implying that the solar cycle is intrinsically non-linear (e.g.
Veselovski \& Tarsina, 2002). Probably the most significant relation
between the cycle shape characteristics is the {\em Waldmeier effect}:
cycles with larger amplitudes are more asymmetric, taking less time to
reach maximum (e.g. Hathaway, \etal, 1994, Li, 1999, Veselovski \&
Tarsina, 2002). That the 11-year oscillation is not harmonic has an
important implication: it introduces infinite number of spectral peaks
(harmonics) with periods $11/n$ years, $n=2,3,...$ (e.g. Polygiannakis
\etal, 1996, Mursula \etal, 1997). All those general features were
successfully described by a simplified mono-parametric Van der Pol
non-linear RLC oscillator model, further shown to accurately reproduce
the observed sunspot record and extended epochs of suppressed activity,
given the maximum of each cycle (Polygiannakis \etal, 1996).

Systematic search for periodicities other than the basic 11-year in
sunspot number, solar activity indices and, consequently, in the
interplanetary, geomagnetic, earth rotation fluctuations and climate
parameters lead to numerous claims, also depending on the nature of the
analyzed parameter (as regarding its sensitivity to the solar activity
periodicities), the periodicity detection method and the finite size of
the records:

(a) Periodicities longer than 11 years, such as 180 years, 90 years
('Gleiseberg cycle'), 45 and 22 years (e.g. Cohen \& Lintz, 1974, Sonnet
1982), perhaps subharmonics (i.e. integer multiples) of the basic
11-year periodicity, and several others (Norderman \& Trivedi, 1992),
intended to describe the long-term modulation of the sunspot cycle
amplitudes as periodic, in contrast to the chaotic or stochastic
theoretical scenarios discussed above.

(b) 11-year periodicity harmonics, namely of 5.5 (FYO, five-year
oscillation) and 3.7 years in sunspot number, other solar activity and
solar-terrestrial connection parameters as the geomagnetic field and
earth rotation disturbances (e.g. Currie, 1976, Courtillot \& Le Mouel,
1976, Suguira, 1980, Carta \etal, 1982, Kondor, 1993, Djurovic \&
Paquet, 1996, Mursula \etal, 1997, Nayar \etal, 2002).

(c) Numerous shorter periodicities of transient character, detected over
one or few 11-year cycles, with most frequently appearing in
bibliography the 2-year (QBO, quasi-biennial) oscillation in sunspot
number (Shapiro \& Ward, 1962, Apostolov, 1985), solar radio flux
(Hughes \& Kesteven, 1981), solar magnetic helicity (Bao \& Zhang,
1998), other solar activity, solar-terrestrial connection and
atmospheric parameters and earth rotation rate (e.g. Chao, 1989,
Djurovic \& Paquet, 1993, Kane, 1997). Oftenly discussed are also
periodicities of 154, 129, 103, 77, 51 days in solar flare occurrence
and sunspot area (Riegel \etal, 1984, Bogard \& Bai, 1985, Dennis, 1985,
Kile \& Cliver, 1991, Bai \& Surrock, 1991, Carbonel \& Ballester,
1992).

The 11-year sunspot (equivalently, 22-year solar magnetic) cycle
phenomenon, is now believed to be due to a magneto-hydrodynamic dynamo
action, periodically regenerating the solar magnetic field at the basis
of the convection zone. The cause and importance of periodicities other
than the 11-year remain so far unclear. The periodicities of 154, 129,
103, 77, 51 days were proposed to be subharmonics of a 25.5 days
fundamental periodicity of some 'clock mechanism' in the sun, modeled by
an oblique rotator, (Bai \& Surrock, 1991, Sturrock \& Bai, 1992).
Alternatively it was proposed that the 154-day periodicity and others
can be related to the normal modes of the solar interior oscillations
(Wolff, 1983, 1992). The source of the QBO may be connected with
periodic poleward streams of magnetic flux, first discussed by Howard
and LaBonde (1981). Based on magnetic helicity observations (Bao \&
Zhang, 1998), it was proposed a {\em double-cycle} solar magnetic dynamo
model, with one dynamo operating at the basis of the convection zone
(11-year cycle) due to radial shear and the other at the top of the
convection zone (with 2-year periodicity) due to latitudal shear
(Benevolenskaya, 1998, 2000). Also supporting this model, further
analysis of the solar magnetic field data showed that the active region
magnetic field exhibits an intense, short-scale component varying with
2-year periodicity (Erofeev, 2001). It was also proposed that the
sunspot number can be well fitted by a superposition of the usual
11-year cycles and wave trains with periodicity continuously varying
from 38 months at solar maximum, to 21 months towards solar minimum
(Krasotkin, 2001).

The sunspot number and all other solar activity parameters also exhibit
non-periodic intense fluctuations that should be attributed to the
episodic character of short-timescale solar phenomena. The solar
magnetic field on the photosphere exhibits complex organization and
sudden, intermittent energy emissions as the solar flares, in all well
modeled as if the solar corona is in a self-organized critical state
(SOC). The corresponding cellular automata models are in good agreement
with the observed power-law distributions of flare released energy and
occurrence rate (e.g. Dennis, 1985, Lu \& Hamilton, 1991, Lu \etal,
1993, Georgoulis \& Vlahos, 1998, Isliker \etal, 2000, 2001,
Anastasiadis, 2002). The emergence of a flare drastically reorganizes
the magnetic fields in solar active regions, hence is expected to affect
the underlying sunspot and active region lifetimes, thus leaving an
imprint of a, characteristic of SOC, $1/f$ tail at high frequencies in
all the solar-activity related parameters, as the here considered
sunspot index.

From the timeseries point of view the sunspot number record provides a
good example of mixture of, both stable and transient, periodicities of
significantly varying amplitudes and of high-frequency, noise-like
components (e.g. Watari, 1996). From the physical point of view it
promises to provide useful information about the, not completely
understood, nature of the solar activity, as connected to the overall
solar magnetic field regeneration mechanism and, at smaller timescales,
the generation of active, transient solar phenomena, such as flares and
coronal mass ejections.

\subsection{Wavelet spectra of sunspot index}

For this application we the record of 2048 recent monthly sunspot number
timeseries extending from June 1831 to February 2002, compiled from the
SIDC RWC Belgium World Data Center for the Sunspot Index catalogues. The
number of points was selected being a power of 2 in order to avoid edge
effects in the fast Fourier transforms used to derive the wavelet
scalograms, thus also omitting the record before 1830 which are of
questionable reliability (e.g. Sonett 1983). The use of monthly-averaged
numbers excludes effects related to the solar differential rotation,
with period of 27 days at the solar equator, and the time-life of each
sunspot, persisting for one or for a few solar rotations.

The wavelet transform was calculated using the Morlet wavelet (equ.
$(\ref{gabor2wavelet})$), filtering the timeseries for each wavelet
scale in the frequency domain for $N=100$ total wavelet periods, sampled
logarithmically between the minimum sampling time (1 month) and the
total timeseries duration (2048 months) and then inverse Fourier
transforming it in the time domain (power theorem, see
$(\ref{powertheorem})$). Note that the total number of period samples,
$N$, used to derive the wavelet scalogram is arbitrary, depending on the
information content in the record (defining the complexity of the
scalogram). Too large $N$ may lead to mode splitting effects and too
small to poorly sampled scalogram. Generally, selecting $N$ being of the
order of $1/10$ of the total number of points in the timeseries (here
2048) gives satisfactory results. For a review on continuous wavelet
transform and the details of the related numerical techniques used to
derive the wavelet scalogram, see Torrence \& Compo (1998). Once the
wavelet scalogram is calculated it is parsed twice, once across time at
each scale and once across scales for each time instance, testing
successive triads of amplitudes and keeping only the locally maximal,
thus obtaining the two skeleton scalograms.

In figure 1(a) we present the monthly-averaged sunspot index. The
well-known 11-year periodicity and the asymmetry of each cycle are
prominent. Also, the fluctuations of the index are larger near each
cycle's maximum.

In figure 1(b) the corresponding wavelet scalogram is shown. Notice the
stable 11-year periodicity, having nearly constant period with time. The
spectrum is rich in higher-frequency components too, and the intensity
of higher frequencies is larger at cycle's maxima, since the fluctuations
are more intense there. Note that there is some arbitrarily in the image
of the scalogram (as this applies to the representation of any wavelet
scalogram and more generally time-frequency distribution), in the sense
that appropriate color or grey scales must be attributed to the range of
wavelet amplitudes, resulting in visually richer or poorer images.

The two skeleton scalograms, extracted from the overall wavelet one, are
shown in figures 1(c) and 1(d). In the locally scale-invariant spectrum
it can be noticed that the lines of maxima proceed towards larger
periods at the cycle's maxima, since they represent the center of each
cycle as an overall structure. The instantly maximal skeleton spectrum
clearly shows the 11-year periodicity as a nearly horizontal line and
others, also quite stable. Periodicities of about 22-years are also
prominent, probably being sub-harmonics of the 11-year one. Near the
period of 1 year the periodicities appear having significant evolution
with time, lasting for many decades (several cycles). The periodicities
of smaller periods seem becoming increasingly unstable and intermittent
(i.e. the lines have more and more gaps), probably due to their small
amplitude, and the presence of self-similar noise-like fluctuations.

The total wavelet power of the overall and the two skeleton spectra is
shown in figure 2. The main spectral peaks are indicated with arrows,
noting the corresponding period in years. On the first column of table 1
we give the period of each of these peaks. The corresponding period
error estimates represent the width between successive wavelet periods,
as sampled to derive the wavelet transform (equ.
$(\ref{gabor2wavelet})$).

The overall wavelet spectrum is poor, exhibiting only few spectral
peaks, namely at the 11, 5.5, 2 and 1 year, and an $1/f$ tail at periods
smaller than 1-year. However, the instantly maximal skeleton spectrum
exhibits several spectral peaks, and this verifies its value: apart from
the dominant 11-year periodicity, 3 of its harmonics, expected due to
the 11-year cycle rise-fall asymmetry, with periods $11/n, \; n=2,3,4$
(i.e. $5.5, 3.67$ and $2.75$ years) and rapidly decreasing amplitudes at
shorter periods are also prominent. The weaker peaks of $16, 13$ and
$8.8$ years seem related with the main 11-year periodicity, probably
being a mode-splitting effect. The long, $28, 24$ and $19$ year
periodicities are probably due to mode-splitting of a 22-year
periodicity, being the main 11-year sub-harmonic, or can be interpreted
as modulation effects of the cycle's amplitudes (see the discussion for
periods longer than 11-years in the previous section).

At shorter periods, a set of 10 spectral peaks with less rapidly
decreasing amplitudes (nearly as $1/f$ with frequency) exists, with
periods given in Table 1. Note that the QBO and the periodicities of
154, 129, 103, 77 days, i.e. 0.42, 0.35, 0.28, 0.21 years (previously
reported in various solar activity parameters, as discussed in the
previous section), are recovered.

As discussed in the previous section, according to the 'clock mechanism'
scenario (Bai \& Surrock, 1991, Sturrock \& Bai, 1992) the 154, 129,
103, 77 day periodicities constitute the $6^{th},5^{th},4^{th}$ and
$3^{th}$ subharmonics of a 25.5 days fundamental periodicity. However
the remaining 6 peaks are not easily explained in a similar way as even
higher subharmonics. The QBO should be the $29^{th}$ subharmonic of the
25.5 day periodicity, however the subharmonics in between are not
observed in the spectrum.

Alternatively, here we propose that all these periodicities are actually
harmonics of the QBO periodicity, which then itself should be
non-harmonic. In table $1$ we present the periodicities found and also
those predicted by the hypothesis of {\em only two non-harmonic
fundamental oscillations, of 11 and 2 years}. Note that $12$ out of $15$
detected spectral peaks, with periods of 11-year and shorter can be
explained solely with this hypothesis. However:

(a) Two of the predicted QBO harmonics, of 0.29 and 0.22 years are not
prominent in the spectrum, possibly being merged, due to noise, with the
expected 0.33 and 0.2 year peaks respectively.

(b) The expected single spectral peak at 1-year period is wide,
extending from about 0.9 to 1.6 years, within which 3 merged peaks of
$0.97,1.07, 1.4$ period are detectable. Possibly this peak widening is
due to evolution of ephemeral, evolving periodicities (as seen in fig.
1(d)), indicative of transient underlying solar activity, or,
alternatively, the $0.97$ and $1.4$ year peaks are additional
periodicities (independent of the QBO) of unknown origin.

\begin{table}

\caption{period (in years) of the spectral peaks in instantly maximal
wavelet skeleton spectrum and those theoretically predicted from
11 and 2 year asymmetric oscillators.}
\begin{flushleft}
\begin{tabular}{cc}
\hline
\noalign{\smallskip}
Observed & Theoretical   \\
\hline
$28 \pm 2$ & \\
$24 \pm 2$ & \\
$19 \pm 1$ & \\
$16 \pm 1$ & \\
$13 \pm 1$ & \\
\hline
     & $11/n$ years \\
\hline
\noalign{\smallskip}
$10.8 \pm 0.8$ & $11$ \\
$8.8 \pm 0.7$ & \\
$5.6 \pm 0.4$ & $5.5$ \\
$3.8 \pm 0.2$ & $3.67$ \\
$2.8 \pm 0.2$ & $2.75$ \\
\hline
\noalign{\smallskip}
     & $2/n$ years \\
\hline
$2.1 \pm 0.1$   & $2$ \\
$1.4 \pm 0.1$ & \\
$1.07 \pm 0.08$ & $1$ \\
$0.97 \pm 0.06$ & \\
$0.68 \pm 0.05$ & $0.67$ \\
$0.49 \pm 0.04$ & $0.5$ \\
$0.42 \pm 0.04$ & $0.4$ \\
$0.34 \pm 0.02$ & $0.33$ \\
    & $0.29$ \\
$0.26 \pm 0.02$ & $0.25$ \\
    & $0.22$ \\
$0.20 \pm 0.01$ & $0.2$ \\
\noalign{\smallskip}
\hline
\end{tabular}

\end{flushleft}
\end{table}

The scale maximal skeleton spectrum is affected only by the main 11-year
periodicity, appearing as a dominant spectral peak, while at timescales
of 1-3 years, where the main QBO peak resides, the scale maximal
spectral power is reduced. At periods smaller than 1 year the scale
maximal spectrum exhibits a plateau of nearly zero-slope. This power
law-behavior at high frequencies is expected from $(\ref{Hurst})$ for
Hurst exponent $h=0$. The noise-like component in the sunspot index thus
exhibits power-law (scale-invariant) properties characterizing the $1/f$
noise, supporting the hypothesis of self-organized criticality (SOC) in
the solar activity, previously proposed for the solar flare occurrence,
as discussed in the previous section.

The observation that all the spectral peaks at periods of 2 years and
shorter, here proposed to be harmonics of the QBO, have amplitudes
decreasing nearly as $1/f$ with frequency (clearly differently from the
11-year cycle, the harmonics of which decay much faster), also emerging
in the same frequency range as the $1/f$ noise-like component (apart
from the 2-year peak itself) indicates a connection of these
oscillations to {\em surface} (photospheric) solar activity phenomena,
such as the solar flares. From theoretical point of view this
observation again supports the model of double solar dynamo
(Benevolenskaya, 1998, 2000) in which the 2-year oscillation is due to
latitudal shear at the {\em top} of the convection zone, hence also the
photosphere. The nature and properties of the surface dynamo producing
both the QBO and intermittent bursts of $1/f$-type (SOC), and the
coupling with the main 11-year dynamo at the basis of the convection
zone is not within the scope of the present article, yet are to our
opinion of great interest for solar physics and deserve further
investigation.

\section{Conclusions}

We proposed the parallel use of two complementary {\em skeleton
spectra}, extracted from the overall continuous wavelet spectrum, that
are useful for studying the existence and evolution of stable or
transient periodicities and noise-like components in timeseries. As an
indicative physical application, we analyzed the monthly averaged sunspot
index. The results showed:

(a) Clearer spectral peaks in the instantly maximal, compared to the
overall, wavelet spectrum, that can be further attributed to the action
of only {\em two non-harmonic oscillators}, namely the well-known
11-year and the 2-year (QBO), supporting the double-cycle solar dynamo
model (Benevolenskaya, 1998, 2000).

(b) The existence of a $1/f$ noise background at timescales of 1 year
and shorter, possibly being an imprint of the previously proposed
self-organized critical state (SOC) of the solar magnetic fields and the
intermittent energy releases, such as solar flares (e.g. Lu \& Hamilton,
1991, Lu \etal, 1993).

We hope that the two wavelet skeleton spectra will be further useful in
various signal processing applications, as data compression, filtering,
data fitting and modeling.

\section*{Acknowledgments}

This research of the author J.Polygiannakis is funded by the Greek
National Foundation for Scholarships (IKY). The author J.Polygiannakis
wishes to thank Dr. V.Agapitou of Queen Mary College of London, School
of Mathematical Sciences, for helpful discussions. We express our thanks
to the University of Athens Research Foundation. We are also grateful to
RWC Belgium World Data Center for making the Sunspot Index catalogues
available to the public.

\newpage

\section*{References}

\small

\begin{description}

\item[] Anastasiadis, A.: 2002, {\em J. Atm. Solar-Terr. Phys.}, {\bf 64}, 481-488.
\item[] Apostolov, E.M.: 1985, {\em Bull. Astron. Inst. Czechosl.}, {\bf 36}, 97-102.
\item[] Argoul, F., Arneodo, A., Grasseau, G., Gagne, Y., Hopfinger, E.J., Frisch, U.: 1989, {\em Nature}, {\bf 338}, 51.
\item[] Arneodo, A., Grasseau, G., Holschneider, M.: 1988, {\em Phys. Rev. Lett.}, {\bf 20}, 2281.
\item[] Arneodo, A., Bacry, E., Muzy, J.F.: 1995, {\em Physica A}, {\bf 213}, 231-75.
\item[] Arneodo, A., d'Aubenton-Carafa, Y., Bacry, E., Graves, P.V., Muzy, J.F., Thermes, C.: 1996, {\em Physica D}, {\bf 96}, 291.
\item[] Arneodo, A., Manneville, S., Muzy, J.F.: 1998, {\em Eur. Phys. J. B}, {\bf 1}, 129.
\item[] Aschwanden, M.J., Kliem, B., Schwarz, U., Kurths, J., Dennis, B.R., Schwarz, R.A.: 1998, {\em Astroph. J.}, {\bf 505}, 941-956.
\item[] Bak, P., Tang, C., Wiesenfeld, K.: 1987, {\em Phys. Rev. Lett.}, {\bf 59}, 381-384.
\item[] Bak, P.: 1996, {\em How nature works, the science of self organized criticality}, Copernicus, Springer-Vergag, New York.
\item[] Bai, T., Surrock, P.A.: 1991, {\em Nature}, {\bf 350}, 141.
\item[] Bao, S., Zhang, H.: 1998, {\em Astroph. J.}, {\bf 496}, L43.
\item[] Ballester, J.L., Oliver, R.: 1999, {\em Astroph. J.}, {\bf 522}, L153-L156.
\item[] Banerjee, D., O'Shea, E., Doyle, J.G.: 2000, {\em Solar Phys.}, {\bf 196}, 63.
\item[] Baumjohann, W., Treumann, R.A., Georgescu, E., Haerendel, G., Fornacon, K.-H., Auster, U.: 1999, {\em Ann. Geophys.}, {\bf 17}, 1528.
\item[] Bendjoya, Ph., Slezak, E., Froeschle: 1991, {\em Astron. Astrophys.}, {\bf 251}, 312.
\item[] Benevolenskaya, E.: 1998, {\em Astrophys. J.}, {\bf 509}, 49.
\item[] Benevolenskaya, E.: 2000, {\em Solar Phys.}, {\bf 191}, 247-255.
\item[] Benzi, R., Biferale, L., Crisanti, A., Paladin, G., Vergassola, M., Vulpiani, A.: 1993, {\em Physica D}, {\bf 65}, 352.
\item[] Beylkin, G., Coifman, R., Rokhlin, V.: 1991, {\em Comm. on Pure and Appl. Math.}, {\bf 44}, 141.
\item[] Bijaoui, A., Slezak, E., Mars, G.: 1993, {\em Wavelets, Fractals and Fourier Transforms}, Oxford University Press, Oxford., p. 213.
\item[] Bocchialini, K., Baudin, F.: 1995, {\em Astron. Astroph.} {\bf 299}, 893.
\item[] Bogard, R.S., Bai, T. : 1985, {\em Astroph. J.}, {\bf 299}, L51.
\item[] Carbonel, M., Ballester, J.L.: 1992, {\em Astron. Astroph.}, {\bf 255}, 350-362.
\item[] Carta, F., Chlistovski, F., Mannara, A., Mazzolenni, F: 1982, {\em Astron. Astroph.}, {\bf 114}, 388.
\item[] Castaing, B., Gagne, Y., Hopfinger, E.J.: 1990, {\em Physica D}, {\bf 46}, 117.
\item[] Chakrabarti, K., Garofalakis, M., Rastogi, R., Shim, K: 2001, {\em VLDB Journ.}, {\bf 10}, 199-223.
\item[] Chao, B.F.: 1989, {\em Science}, {\bf 243}, 923-925.
\item[] Cohen, T.J., Lintz, P.R.: 1974, {\em Nature}, {\bf 250}, 398.
\item[] Cohen, A., Kovacevic, J.: 1996, {\em Proc. IEEE}, {\bf 84}, 4, 514.
\item[] Courtillot, V., Le Mouel, J.-L.: 1976, {\em J. Geophys. Res.}, {\bf 81} (17), 2941.
\item[] Currie, R.G.: 1976, {\em Astroph. Space Sci.}, {\bf 39}, 251.
\item[] De Moortel, Ireland, J., Walsh, R.W.: 2000, {\em Astron. Astroph.}, {\bf 355}, L23.
\item[] De Moortel, Hood, A.W., Ireland, J. : 2002a, {\em Astron. Astroph.}, {\bf 381}, 311-323.
\item[] De Moortel, Ireland, J., Hood, A.W., Walsh, R.W. : 2002b, {\em Astron. Astroph.}, {\bf 387}, L13-L16.
\item[] Dennis, B.R.: 1985, {\em Solar Phys.}, {\bf 100}, 65.
\item[] Djurovic, D., Paquet, P.: 1993, {\em Astron. Astroph.}, {\bf 277}, 669-676.
\item[] Djurovic, D., Paquet, P.: 1996, {\em Solar Phys.}, {\bf 167}, 427.
\item[] Eddy, J.A. : 1977, {\em The Ancient Sun}, ed. Pepin,R.A., Eddy, J.A., Merrill, R.B., Pergamon, New York.
\item[] Erofeev, D.V.: 2001, {\em Solar Phys.}, {\bf 198}, 31-50.
\item[] Fligge, M., Solanki, S.K.: 1997, {\em Astron. Astroph. Supl.}, {\bf 124}, 579.
\item[] Fligge, M., Solanki, S.K., Beer, J.: 1999, {\em Astron. Astroph.}, {\bf 346}, 313.
\item[] Fludra, A. : 2001, {\em Astron. Astroph.}, {\bf 368}, 639-651.
\item[] Foster, G.: 1996, {\em Astron. J.}, {\bf 112} (4), 1709.
\item[] Frick, P., Galyagin, D., Hoyt, D.V., Nesme-Ribes, E., Schatten, K.H., Sokoloff, D., Zakharov, V.: 1997, {\em Astron. Astroph.}, {\bf 328}, 670.
\item[] Frick, P., Baliunas, S.L., Galyagin, D., Sokoloff, D., Soon, W.: 1997, {\em Astroph. J.} {\bf 483}, 426.
\item[] Gabor, D.: 1946, {\em J. Inst. Electr. Engin.}, {\bf 93}, 429.
\item[] Georgoulis, M.K., Vlahos, L: 1998, {\em Astron. Astroph.}, {\bf 336}, 721-735.
\item[] Girardi, M., Fadda, D., Escalera, E.; Giuricin, G., Mardirossian, F.; Mezzetti, M.: 1997, {\em Astroph. J.}, {\bf 490}, 56.
\item[] Gimenez de Castro, C.G., Raulin, J.P., Mandrini, C.H., Kaufmann, P., Magun, A.: 2001, {\em Astron. Astroph.}, {\bf 366}, 317-325.
\item[] Hathaway, D.H., Wislon, R.M., Reichmann, E.J.: 1994, {\em Solar Phys.}, {\bf 151}, 177.
\item[] Haynes, P.H., Norton, W.A.: 1993, {\em Wavelets, Fractals and Fourier Transforms}, Oxford University Press, Oxford., p. 229.
\item[] Hempelmann, A., Donahue, R.A.: 1997, {\em Astron. Astroph.}, {\bf 322}, 835-840.
\item[] Hempelmann, A. : 2002, {\em Astron. Astroph.}, {\bf 388}, 540-545.
\item[] Howard, R., LaBonde, B.J.: 1981, {\em Solar Phys.}, {\bf 74}, 131.
\item[] Hoyng, P.: 1988, {\em Astroph. J.}, {\bf 332}, 857.
\item[] Hoyng, P.: 1993, {\em Astron. Astroph.}, {\bf 272}, 321-339.
\item[] Hughes, V.A., Kesteven, M.J.L.: 1981, {\em Solar Phys.}, {\bf 71}, 259.
\item[] Norderman, D.J.R., Trivedi, N.B.: 1992, {\em Solar Phys.}, {\bf 142}, 411-414.
\item[] Hunt, J.C.R., Kevlahan, N.K.R., Vassilicos, J.C., Farge, M.: 1993, {\em Wavelets, Fractals and Fourier Transforms}, Oxford University Press, Oxford.
\item[] Isliker, H., Anastasiadis, A., Vlahos, L.: 2000, {\em Astron. Astroph.}, 363, 1134-1144.
\item[] Isliker, H., Anastasiadis, A., Vlahos, L.: 2001, {\em Astron. Astroph.}, 377, 1068-1080.
\item[] Ireland, J., Walsh, R.W., Harrison, R.A., Priest, E.R.: 1999, {\em Astron. Astroph.}, {\bf 347}, 355-365.
\item[] Kaiser, G.: 1994, {\em A friendly guide to wavelets}, Birkhauser, Boston.
\item[] Kane, R.P.: 1997, {\em Ann. Geophys.}, {\bf 15}, 1581-1594.
\item[] Kile, J.N, Cliver, E.W.: 1991, {\em Asroph. J}, {\bf 370}, 442-448.
\item[] K$\ddot{u}$ker, M., Arlt, R., R$\ddot{u}$diger, G.: 1999, {\em Astron. Astroph.}, {\bf 343}, 977-982.
\item[] Komm, R.W., Gu, Y., Hill, F., Stark, P.B., Fodor, I.K.: 1999, {\em Astroph. J.}, {\bf 519}, 407.
\item[] Kondor, N.N.: 1993, {\em Adv. Space Res.}, {\bf 13} (9), 417.
\item[] Krasotkin, S.A.: 2001, {\em SOLSPA 2001 Euroconf.: Solar and Space weather}, Proc. ESA SP-477.
\item[] Lawrence, J.K., Cadavid, A.C., Ruzmaikin, A.: 2001, {\em Solar Phys.}, {\bf 202}, 27-39.
\item[] Li, K.: 1999, {\em Astron. Astroph.}, {\bf 345}, 1006-1010.
\item[] Lima Neto, G. B., Pislar, V., Durret, F., Gerbal, D.; Slezak, E.: 1997, {\em Astron. Astroph.}, {\bf 327}, 81-89.
\item[] Louis, A.K., Maa$\beta$, P., Rieder, A.: 1997, {\em Wavelets, Theory and Applicatons}, Wiley \& Sons, Chichester, UK., p. 16.
\item[] Lu, E.T., Hamilton, J.R.: 1991, {\em Astroph. J.}, {\bf 380}, L89.
\item[] Lu, E.T., Hamilton, J.R., Mc Tiernan, J.M., Bromund, K.R.: 1993, {\em Astroph. J.}, {\bf 412}, 841.
\item[] Lucek, E.A., Balogh, A.: 1997, {\em Geophys. Res. Lett.}, {\bf 24}, 18, 2387.
\item[] Mallat, S. G.: 1989, {\em IEEE Trans. on pattern analysis and machine intelligence}, {\bf 11}, 7, 674.
\item[] Mallat, S.G.: 1999, {\em A wavelet tour of signal processing}, Second Edition, Academic Press, CA.
\item[] Marsh, M. S., Walsh, R. W., Bromage, B. J. I.: 2002, {\em Astron. Astroph.}, {\bf 393}, 649-659.
\item[] Meszarosova, H., Jiricka, K., Karlicky, M.: 1999, {\em Astron. Astroph.}, {\bf 348}, 1005.
\item[] Muhlmann, W., Hanslmeier, A.: 1996, {\em Solar Phys.}, {\bf 166}, 445.
\item[] Monin, A.S, Yaglom, A.M.: 1975, {\em Statistical fluid mechanics: mechanics of turbulence}, vol. 2, Lumley, J.L. (ed.), MIT Press, Cambridge, Mass.
\item[] Mursula, K.; Usoskin, I.; Zieger, B., 1997, {\em Solar Phys.}, {\bf 176}, 1, 201.
\item[] Mordvinov, A.V., Kuklin, G.V.: 1999, {\em Solar Phys.}, {\bf 187}, 223-226.
\item[] Mordvinov, A.V., Plyusnina, L.A.: 2000, {\em Solar Phys.}, {\bf 197}, 1.
\item[] Morlet, J.G., Arehs, I, Fourgeau, Giard, D.: 1982, {\em Geophysics}, {\bf 47}, 203.
\item[] Mouri, H., Takaoka, M., Kubotani, H.: 1999, {\em Phys. Lett. A}, {\bf 461}, 82.
\item[] Muzy, J.F., Bacry, E., Arneodo, A.: 1991, {\em Phys. Rev. Lett.}, {\bf 67}, 25, 3515.
\item[] Muzy, J.F., Bacry, E., Arneodo, A.: 1993, {\em Phys. Rev. E}, {\bf 47}, 2, 875.
\item[] Nayar Prabhakaran, S.R., Radhika, V.N., Revathy, K., Ramadas, V.: 2002, {\em Solar Phys.}, {\bf 208}, 359-373.
\item[] O'Shea, E., Banerjee, D., Doyle, J.G., Fleck, B., Murtagh, F.: 2001, {\em Astron. Astroph.}, {\bf 368}, 1065.
\item[] O'Shea, E., Murtagh, F., Fleck, B.: 2002, {\em Astron. Astroph.}, {\bf 387}, 642-664.
\item[] Ossendrijver, A.J.H., Hoyng, P.: 1996, {\em Astron. Astroph.}, {\bf 313}, 959-970.
\item[] Otazu, X., Ribo, M., Peracaula, M., Paredes, J.M., Nunez, J.: 2002, {\em Mon. Not. R. Astron. Soc.}, {\bf 333}, 365-372.
\item[] Pantin, E., Starck, J.L.: 1996, {\em Astron. Astroph. Suppl.}, {\bf 118}, 575.
\item[] Polygiannakis, J.M., Moussas, X.: 1996, {\em Solar Phys.}, {\bf 163}, 193-203.
\item[] Press, W.H., Flannery, B.P., Teukolski, S.A., Vetterling, W.T.: 1992, {\em Numerical Recipes}, Cambridge University Press, Cambridge, p. 420.
\item[] Ribes, J.C., Nesme-Ribes, E.: 1993, {\em Astron. Astroph.}, {\bf 276}, 549-563.
\item[] Riegel, E., Share, G.H., Forrest, D.J., Kanbach, G., Reppin, C., Chupp, E.L.: 1984, {\em Nature}, {\bf 312}, 623-625.
\item[] Rigozo, N.R., Echer, E., Vieira, L.E.A., Nordemann, D.J.R. : 2001, {\em Solar Phys.}, {\bf 203}, 179-191.
\item[] Roux, S., Muzy, J.F., Arneodo, A.: 1999, {\em Eur. Phys. J. B}, {\bf 8}, 301.
\item[] Rusmaikin, A.A.: 1983, {\em Stellar and Planetaty magnetism}, Soward, A.M., (ed.), Gordon \& Breach, New York, p. 151.
\item[] Ruzmaikin, A.A., Lyannaya, I.P., Styashkin, V.A., Yeroshenko, E.: 1993, {\em J. Geophys. Res.}, {\bf 98}, 13303.
\item[] Saito, N., Beylkin, G.: 1993, {\em IEEE Trans. on Signal Proc.}, {\bf 41}, 12, 3585.
\item[] Schmitt, D., Schussler, M., Ferriz-Mas: 1996, {\em Astron. Astroph.}, {\em 311}, L1-L4.
\item[] Schwarz, U., Kurths, J., Kliem, B., Kruger, A., Urpo, S.: 1998, {\em Astron. Astroph. Suppl.}, {\bf 127}, 309.
\item[] Sello, S.: 2000, {\em Astron. Astroph.}, 363, 311-315.
\item[] Schneider, K., Kevlahan, N.K.-R., Farge, M.: 1997, {\em Theor. Comp. Fluid Dyn.}, {\bf 9}. 191-206.
\item[] Shapiro, R., Ward, F.: 1962, {\em J. Atmos. Sci.}, {\bf 19}, 127.
\item[] Sonett, C.P. : 1983, {\em J. Geoph. Res.}, {\bf 88}, 3225.
\item[] Soon, W., Frick, P., Baliunas, S.: 1999, {\em Astroph. J.}, {\bf 510}, L135-L138.
\item[] Sturrock, P., Bai, T.: 1992, {\em Astroph. J.}, {\bf 397}, 337.
\item[] Torrence, C., Compo, G.P.: 1998, {\em Bull. Am. Met. Soc.}, {\bf 79}, 61.
\item[] Veselovski, I.S., Tarsina, M.V.: 2002, {\em Adv. Space Res.}, {\bf 29}, 3, 417-420.
\item[] Waldmeier, M. : 1961, {\em The sunspot activity in the years 1610-1960}, Zurich Schulthess and Company AG.
\item[] Watari, S.: 1996, {\em Solar Phys.}, {\bf 168}, 413-422.
\item[] Wolff, C.L.: 1983, {\em Astroph. J.}, {\bf 264}, 667.
\item[] Wolff, C.L.: 1992, {\em Solar Phys.}, {\bf 142}, 187.
\item[] Zhang, Y., Paulson, K.V.: 1997, {\em Pure Appl. Geophys.}, {\bf 149}, 405.
\end{description}

\normalsize

\newpage

\section*{Figure captions}

$\; \; \; \;$ {\bf Figure 1.} The monthly-averaged sunspot index. (b) The corresponding
wavelet scalogram in grey color scale (whiter color corresponds to
larger wavelet amplitudes). (c) The scale maximal and (d) the instantly
maximal wavelet skeleton spectra.

\vspace{1cm}

{\bf figure 2.} The total wavelet power of the overall (upper curve), the
instantly maximal (middle curve, the period in years of the spectral
peaks is also noted) and scale maximal wavelet skeleton spectra in
common, arbitrary units.

\end{document}